\documentclass{aa}
\begin{document}

\title{On the age of PSR B\,1509-58}

\author{V.V.Gvaramadze\inst{1,2}\thanks{{\it Address for
correspondence}: Krasin str. 19, ap. 81, Moscow, 123056, Russia
(vgvaram@mx.iki.rssi.ru)}}

\institute{Abastumani Astrophysical Observatory, Georgian Academy of
Sciences, A.Kazbegi ave. 2-a, Tbilisi, 380060, Georgia
\and Sternberg State Astronomical Institute, Universitetskij Prospect 13,
Moscow, 119899, Russia
}

\date{Received 26 April 1999 / accepted 17 May 2001}

\abstract{
It is generally accepted that the
PSR B\,1509-58 is associated with the supernova remnant (SNR)
MSH\,15-52 (G\,320.4-01.2). The spin-down age of the pulsar is $\simeq 1700$
years, while the size and the general appearance of the SNR suggest
that this system is much older. A few possible explanations of
this discrepancy have been put forward.
We offer an alternative one and suggest that the high spin-down
rate of the pulsar characterizes only a
relatively short period of its (present) spin
history, and that the enhanced braking torque is connected with
the interaction between the pulsar's magnetosphere and the dense matter
of a circumstellar clump (created during the late evolutionary
stages of the supernova (SN) progenitor star). Our suggestion
implies that the ``true" age of PSR B\,1509-58 could be much
larger than the spin-down age, and therefore the SNR
MSH\,15-52 is a middle-aged remnant similar to the Vela SNR
(G\,263.9-3.3). We also suggest that
the dense (neutral) gas of the circumstellar clump could be responsible
for the enhanced neutral hydrogen absorption towards PSR B\,1509-58,
and that the optical emission of an optical
counterpart for PSR B\,1509-58 should rather be attributed to a
bow shock around this pulsar than to the pulsar itself.
\keywords{Stars: neutron -- pulsars: individual: B\,1509-58 -- ISM:
          bubbles -- ISM: individual objects: MSH\,15-52 --
          ISM: supernova remnants}
         }

\maketitle

\section{Introduction}
%
\object{PSR B\,1509-58} (\cite{s2}, \cite{m2}) is situated not far
from the geometrical centre of the SNR \object{MSH\,15-52} (or
\object{G 320.4-01.2}), and their association is beyond any doubt.
However, this association causes a number of difficulties for the
understanding and interpretation of the observational data. The
problem is that the size and the general appearance of the SNR
suggest that it should be much older than it follows from the
pulsar age estimates.

It is usually assumed that the rotational frequency $\Omega$ of a
pulsar decreases according to the relation ${\dot \Omega} \,=\,
-\,K\Omega ^n $, where $K$ depends upon the physics of the
slow-down mechanism, and $n = \Omega {\ddot \Omega}/{\dot
\Omega}^2$ is the braking index. Assuming constant $K$ and $n$,
and provided that the initial spin period $P_{\rm i}$ of the
pulsar was much smaller than the current period $P=2\pi /\Omega$,
one can estimate the characteristic spin-down age $\tau \,=\,
P/(n-1){\dot P}$. For $P\simeq 0.15$ s, ${\dot P} \simeq 1.49
\cdot 10^{-12} \, {\rm s\,s}^{-1}$ and $n=2.84$ (\cite{w},
\cite{mdn}, Kaspi et al. \cite{kas}), one derives an age of PSR
B\,1509-58 of $\simeq 1700$ years, i.e. it is nearly as young as
the \object{Crab pulsar}. The spin-down age could be even less by
a factor of $[1-(P_{\rm i} /P)^{n-1}]^{-1}$, if the pulsar was
born with $P_{\rm i}$ of $\simeq 0.1$ s (see e.g. \cite{spr}).
These estimates are at odds with the age estimates for MSH\,15-52
(Seward et al. \cite{s3}, \cite{van}, \cite{kam}), which show that
the SNR is a much older object.

To reconcile the ages of the pulsar and the SNR, Seward et al.
(\cite{s3}) considered two possibilities: 1) MSH\,15-52 is a young
SNR, and 2) PSR B\,1509-58 is an old pulsar. The first one implies
(in the framework of the Sedov-Taylor model) that the SN explosion
was very energetic and occured in a tenuous medium (see also
\cite{bh}). This point of view is generally accepted (e.g. Kaspi
et al. \cite{kas}, \cite{gre}, \cite{tru}, Gaenzler et al.
\cite{gae}). The second possibility implies that $\tau$ is at
least few times shorter than the ``true" age (Seward et al.
\cite{s3}). This possibility was re-examinated by \cite{bl}.
Assuming that the pulsar spin-down is mostly due to the
electromagnetic torque, they suggested that the torque grew within
the last $\simeq 10^3$ years due to the growth of the pulsar's
magnetic field (see also \cite{mu}). In this case, the coefficient
$K$ is an increasing function of time and therefore the ``true"
age of the pulsar could be as large as follows from the age
estimates for the SNR. In this paper we offer an alternative
explanation for the increase of the braking torque (Sect. 2), viz.
we suggest that it could be episodically enhanced due to the
interaction of the pulsar's magnetosphere with dense clumps of
circumstellar matter (Sect. 3). Sect. 4 deals with some issues
related to our suggestion.

\section{The spin-down of PSR B\,1509-58}
It is known that the electromagnetic torque acting on a rotating,
magnetized body (e.g. a neutron star) immersed in a plasma is
enhanced as compared with the torque in vacuum (\cite{gin}, see
also Istomin \cite{i}). It was mentioned by Istomin (\cite{i})
that for the increase of the slow-down torque of a pulsar it is
suffient to have a dense plasma in the vicinity of the light
cylinder, since just in this region the pulsar loses its
rotational energy due to the acceleration of particles. The
particles of the ambient medium penetrating into the region of the
light cylinder are accelerated there to velocities comparable with
the speed of light and then leave this region (Istomin \cite{i}).
This presumably equatorial outflow (cf. \cite{bal}, \cite{kin})
carries away the pulsar's angular momentum and is responsible for
the enhanced braking of the pulsar. We suggest (see also
\cite{g9a}; cf. \cite{y}, \cite{ik}) that just this effect is
responsible for the present high spin-down rate of PSR B\,1509-58,
i.e. that the pulsar loses its rotational energy mainly due to the
acceleration of protons of the ambient medium arriving at the
light surface at the rate $\dot{M}$:
\begin{equation}
\label{1}
|{\dot E}| \, = \, |I\Omega \dot{\Omega}| \, = \, \gamma _{\rm p}
\dot{M} c^2  \, \, ,
\end{equation}
where $I \simeq 10^{45} \, {\rm g}\,{\rm cm}^2$ is the moment
of inertia of the pulsar,
$c$ is the speed of light, and $\gamma _{\rm p}$ is the Lorentz factor of
accelerated protons. For $\Omega = 41.89 \, {\rm s}^{-1}$ and
$\dot{\Omega} = -4.25\cdot 10^{-10} \, {\rm s}^{-2}$
(e.g. Kaspi et al. \cite{kas}), and $\gamma _{\rm p} \simeq 1$
(Istomin \cite{i}),
one has $\dot{M} \simeq 2.2\cdot 10^{16} \, {\rm g}\,{\rm s}^{-1}$.
We also suggest that the pulsar moves through the
inhomogeneous ambient medium and episodically plunges into clumps of
dense matter. If the density of clumps is sufficiently high
(see Sect. 3), one can expect a temporary
increase of $\dot{\Omega}$ every time the pulsar
travels through a clump (we believe that just this situation
takes place now). The corresponding spin-down age will be
less than estimated when the pulsar moves
through the low-density interclump medium.

It is clear that the presence of radio emission of the pulsar
means that the ambient medium does not penetrate far beyond the
light surface. Assuming that the ram pressure of the accreting
medium is equal to the magnetic pressure at the light surface (cf.
\cite{kin}), one has an estimate of the surface magnetic field of
the pulsar\footnote{Note that this estimate is about ten times
smaller than what follows from the usual dipole formula
(\cite{w}).}:
\begin{displaymath}
B_{\ast} \, = \, \left[2(2GM_{\ast})^{1/2}
\dot{M} r_{\rm L}^{7/2} /r_{\ast} ^6
\right]^{1/2}  \, \simeq 2.9 \cdot 10^{12} \, {\rm G} \, \, ,
\end{displaymath}
where $G$ is the gravitational constant, $M_{\ast}=1.4 \,
M_{\odot}$ and $r_{\ast} =10^6$ cm are the
mass and the radius of the pulsar, and $r_{\rm L} = c/\Omega
\simeq 7.1\cdot 10^8$ cm is the radius of the light cylinder;
for simplicity we assumed that the pulsar magnetic field is dipolar.
Given this value of $B_{\ast}$, one can estimate the ``vacuum"
values of $\dot{\Omega}$ and $\tau$: $\dot{\Omega} _0 = -2\mu ^2 \Omega
^3 /3c^3 I \simeq -1.36\cdot 10^{-11} \, {\rm s}^{-2}$,
where $\mu = B_{\ast} r_{\ast}^3$ is the magnetic momentum
of the pulsar, and $\tau _0 \leq - \Omega/2\dot{\Omega} _0 \simeq
4.9\cdot 10^4 \, {\rm yr} \simeq 30\tau$ (here we assumed that the
``vacuum" braking index is equal to 3). This means that the ``true"
age of the pulsar could be as large as $\tau _0$ (provided that
$P_{\rm i} \ll P$)
and that the SNR MSH\,15-52 could be a middle-aged
remnant similar to the \object{Vela SNR} (\object{G\,263.9-3.3}).

\section{Dense clumps in the central part of SNR MSH\,15-52}

We suggest that PSR B\,1509-58 and SNR MSH\,15-52 are the remnants
of the SN explosion of a massive star (\cite{g9b}). In this case,
the structure of MSH\,15-52 could be determined by the interaction
of the SN blast  wave with the ambient medium reprocessed by the
joint action of the ionizing emission and stellar wind of the SN
progenitor star (\cite{mck}, \cite{shu}, \cite{cio}, \cite{cl},
\cite{f}, \cite{d}, \cite{g9b},\cite{g9c}, \cite{g0a}). The outer
shell of the SNR could arise due to the abrupt deceleration of the
SN blast wave after it encounters the density jump at the edge of
the bubble created by the fast stellar wind during the
main-sequence or the Wolf-Rayet (WR) stages. On the other hand,
some structures in the central part of the SNR\footnote{Note that
these structures could be significantly offset from the
geometrical centre of a middle-aged SNR due to the proper motion
of the SN progenitor star (\cite{g9b}, \cite{g0a},\cite{g0b}).}
could be attributed to the interaction of the SN blast wave with
the circumstellar material lost during the late evolutionary
stages of the SN progenitor star (this is the material that
determines the appearance of young type\,II SNRs, e.g. SN\,1987A
(e.g. \cite{mcc}) or Cas\,A (e.g. \cite{gar}, \cite{bo})). During
the red supergiant (RSG) stage a massive star loses a significant
part (about two thirds) of its mass in the form of a slow, dense
wind. This matter occupies a compact region with a characteristic
radius of few parsecs (the high-pressure gas in the main-sequence
bubble significantly affects the spreading of this region
(\cite{ce}, \cite{d})). Before the SN exploded, the progenitor
star (of mass $> 15-20 \, M_{\odot}$) becomes for a short time a
WR star (e.g. \cite{Van}). At this stage, the fast stellar wind
sweeps up the slow RSG wind and creates a low-density cavity
surrounded by a shell of swept-up circumstellar matter. The shell
expands with a nearly constant velocity $v_{\rm sh} \simeq
(\dot{M} _{\rm WR} v_{\rm WR} ^2 v_{\rm RSG}/3\dot{M} _{\rm
RSG})^{1/3}$, where $\dot{M} _{\rm WR} , \dot{M} _{\rm RSG}$ and
$v_{\rm WR}, v_{\rm RSG}$ are, correspondingly, the mass-loss
rates and wind velocities during the WR and RSG stages  (e.g.
\cite{dy}),  until it catches up the shell separating the RSG wind
from the main-sequence bubble. For parameters typical for RSG and
WR winds, one has $v_{\rm sh} \simeq 100-200 \, {\rm km}\,{\rm
s}^{-1}$. The interaction of two circumstellar shells results in
Rayleigh-Taylor and other dynamical instabilities, whose
development is accompanied by the formation of dense clumps moving
with radial velocities of $\simeq v_{\rm sh}$ (\cite{gar}). The
dense clumps could originate much closer to the SN progenitor star
due to the stellar wind acceleration during the transition from
the RSG to the WR stage (\cite{br}). The number density of clumps
is $\geq 10^5 \, {\rm cm}^{-3}$ provided they are not fully
ionized and were able to cool to a temperature of $\leq 10^2$ K
(\cite{br}). Direct evidence of the existence of high-density
clumps close to the SN explosion sites follows from observations
of young SNRs. For example, the optically emitting gas of
quasi-stationary flocculi in \object{Cas\,A} is characterized by a
density of $\simeq 10^4 \, {\rm cm}^{-3}$ and a temperature of
$\geq 10^4$ K (e.g. \cite{l}). Assuming that the optical emission
of a floccule comes from an ionized ``atmosphere" around the
neutral core, one can estimate the density of the core to be $\geq
10^6 \, {\rm cm}^{-3}$, provided that the temperature of the core
is $\leq 10^2$ K. Similar estimates could also be derived from
observations of the optical ring around \object{SN\,1987A}, the
inner ionized ``skin" of which has nearly the same parameters
(e.g. \cite{pl}) as the optically emitting gas of flocculi in
Cas\,A, or from observations of some other young SNRs (e.g.
\cite{chu}, \cite{cd}). The radial velocity of flocculi in Cas\,A
ranges from $\simeq 80$ to $\simeq 400 \, {\rm km}\,{\rm s}^{-1}$
(e.g. \cite{l}).

Initially, the new-born pulsar moves through the low-density
cavity created by the fast wind of the presupernova star until it
plunges into the first dense clump on its way. This happens at the
moment $t \sim r_{\rm cav} /v_\ast$, where $r_{\rm cav} \simeq
1-2$ pc is the radius of the cavity, $v_\ast = v_{\rm p} - v_{\rm
cl}$, $v_{\rm p}$ and $v_{\rm cl}$ are respectively the velocities
of the pulsar and the clump. For $v_{\rm p} \simeq 150 \, {\rm
km}\,{\rm s}^{-1}$ and $v_{\rm cl} \simeq 100 \, {\rm km}\,{\rm
s}^{-1}$, one has $t\simeq (2-4)\cdot 10^4$ years\footnote{No
proper motion for PSR B\,1509-58 has yet been detected (Kaspi et
al. \cite{kas}, Gaensler et al. \cite{gae}), though the recent
finding of the candidate optical counterpart of this pulsar
(Caraveo et al. \cite{ca}, see also \cite{mi} and Sect.4) allows
one to hope that it will be detected in due course. For the
distance to the pulsar of $5d_5$ kpc (e.g. Gaensler et al.
\cite{gae}), a velocity of $150\, {\rm km} \, {\rm s}^{-1}$
corresponds to a proper motion of $\simeq 6 \, {\rm mas}\,{\rm
yr}^{-1}$; this is much smaller than the upper limits on the
pulsar's proper motion given by Kaspi et al. (\cite{kas}) and
Gaensler et al. (\cite{gae}).}. Let us assume that all matter
captured inside the accretion radius $r_{\rm acc} =2GM_{\ast}
/v_{\ast} ^2$ ($v_{\ast} \gg c_{\rm s}$, where $c_{\rm s}$ is the
sound speed in the cold, dense circumstellar clump) of the pulsar
moving through the clump penetrates into the region of the light
cylinder, where it is accelerated to relativistic velocities and
then leaves this region in the form of equatorial outflow (cf.
Istomin \cite{i} and \cite{kin}). The rate at which the ambient
medium arrives at the light cylinder could be estimated as
\begin{equation}
\label{2}
\dot{M} \, = \, \pi r_{\rm acc} ^2 n_{\rm cl} m_{\rm p} v_{\ast} \, \, ,
\end{equation}
where $n_{\rm cl}$ is the number density of the clump, and $m_{\rm p}$ is
the mass of a proton.
If the pulsar braking is indeed mainly due to the
acceleration of circumstellar
protons arriving at the light surface, then one obtains from (\ref{1})
and (\ref{2}) that
$n_{\rm cl} \simeq 3.8\cdot10^6 \, v_{\ast ,50} ^2 \, {\rm cm}^{-3}$,
where $v_{\ast ,50} =v_{\ast} /50 \, {\rm km}\,{\rm s}^{-1}$.
The time it takes for PSR B\,1509-58 to cross the clump, $t^{'} \,
\simeq \, l_{\rm cl} /v_{\ast}$, where $l_{\rm cl}$ is the characteristic
size of clumps, should be larger than the time since
the pulsar discovery, i.e. $t^{'} > 30$ years. This
requirement results in $l_{\rm cl} \geq 5\cdot10^{15} v_{\ast ,50}$ cm
and $M_{\rm cl} > 0.0002 v_{\ast ,50} ^3
M_{\odot}$, where $M_{\rm cl}$ is the
characteristic mass of clumps. For the SN
progenitor star of mass $\geq 15M_{\odot}$, the mass of the circumstellar gas
(i.e. the matter lost during the RSG stage) is about $10M_{\odot}$, and
the number of clumps is $<5\cdot10^4$. For the current radius
of the region occupied by circumstellar matter of about 10 pc
(\cite{g9b}), one finds a covering factor of the clumpy
circumstellar material (i.e. the fraction of the sphere occupied by
clumps) $\simeq 10^{-4}$.

For accretion to occur, the standoff radius $r_{\rm s}$ of the bow
shock (formed by the outflow of relativistic particles) should be
less than $r_{\rm acc}$. For the spherically symmetric outflow,
one has $r_{\rm s} =(\alpha |{\dot E}|/4\pi n_{\rm cl}m_{\rm p}
cv_{\ast}^2)^{1/2} < r_{\rm acc} =2GM_{\ast} /v_{\ast} ^2$, where
we assume that only a fraction $\alpha < 1$ of the spin-down
luminosity $|{\dot E}|$ is transferred to the ambient medium (cf.
\cite{koc}). This condition can be re-written as $\alpha <
(\gamma_{\rm p}/4)^{-1} v_{\ast} /c \simeq 10^{-3} \gamma_{\rm p}
^{-1} v_{\ast ,50}$ (cf. \cite{koc}, \cite{m5}), i.e. $\alpha$
should be much smaller than the usually adopted value of $\simeq
1$ (e.g. \cite{kul}, \cite{co}). Weak coupling ($\alpha \ll 1$) of
the pulsar wind with the ambient medium is consistent with an
outflow composed of highly relativistic particles (e.g. \cite{koc}
and references therein). Alternatively, if the outflow of
relativistic particles is confined to the vicinity of the
rotational equatorial plane, one can expect that the ambient
matter accretes onto the pulsar's magnetosphere along the polar
directions. Another possibility is that the ambient matter
penetrates in the pulsar wind bubble through instabilities in the
bow shock front. In the latter both cases $r_{\rm s}$ could be
larger than $r_{\rm acc}$, and one can adopt $\alpha \simeq 1$
(see next section).

\section{Discussion}

We now discuss some consequences of our proposal that the braking of
PSR B\,1509-58 is mostly due to the interaction of the pulsar's
magnetosphere with the dense matter of a circumstellar clump.

First, we consider the contribution of the circumstellar matter to the
neutral hydrogen absorption toward the pulsar.
The low covering factor of clumps (see Sect. 3)
implies that this contribution is
\begin{equation}
\label{3}
N_{\rm cl} \, = \, \int_{r_{\rm L}}^{r_{\rm acc}} \limits
n_{\rm ff} (r)dr +
n_{\rm cl} (r_{\rm cl} -r_{\rm acc}) \, \, ,
\end{equation}
where $n_{\rm ff} (r)=\dot{M}/4\pi m_{\rm p} r^2 v_{\rm ff}$ is the number
density of the infalling gas captured inside the accretion radius,
$v_{\rm ff} =(2GM_{\ast}/r)^{1/2}$ is the free-fall velocity,
$r_{\rm cl}$ is the line of sight thickness of the dense neutral gas
located between the pulsar and the observer ($r_{\rm cl} \leq l_{\rm cl}$).

It was pointed out by Strom (\cite{st}) that a ROSAT observation
of the SNR MSH\,15-52 indicates a larger absorption towards PSR
B\,1509\,-\,58 than seen from the bright northwest part of the
SNR's shell (known as \object{RCW\,89}). The comparison of neutral
hydrogen absorption data (see \cite{gre}, \cite{tru}, \cite{tam},
\cite{ma}, \cite{ro}) shows that the excess of absorption towards
the pulsar could be as large as $\simeq (1-5)\cdot 10^{22} \, {\rm
cm}^{-2}$. This discrepancy could be interpreted as a sign that
the pulsar is more distant than the SNR, and therefore that these
two objects are not physically associated with each other (Strom
\cite{st}). It is also possible that ``the spectral analysis is
not detailed enough to provide the correct parameters"
(\cite{tru}). Another possibility is that the H\,I column density
distribution is really inhomogeneous across the SNR (cf.
\cite{tru}). We favour the last possibility and suggest that the
excess of absorption towards PSR B\,1509-58 is due to the dense
neutral gas around the pulsar. One can use Eq. (\ref{3}) to set an
upper limit on $r_{\rm cl}$. For the parameters adopted above, and
assuming that $N_{\rm cl} \leq (1-5)\cdot 10^{22} \, {\rm
cm}^{-2}$, one has $r_{\rm cl} \leq (0.2-1.2)\cdot10^{16} v_{\ast,
50} ^{-3}$ cm. This estimate shows that if our explanation of the
age discrepancy is correct, then one might expect that in the near
future (i.e. after a lapse of $\simeq r_{\rm cl}/v_{\ast,
\parallel}$, where $v_{\ast,
\parallel}$ is the line of sight component of $v_{\ast}$) the
first derivative of the pulsar's spin period will suffer a
significant decrease.

Second, let us discuss the candidate optical counterpart for PSR
B\,1509-58 proposed by Caraveo et al. (\cite{ca}). Caraveo et al.
pointed out that the luminosity of the optical counterpart
($L_{\rm V} \sim 6.5\cdot 10^{32} d_5 ^2 \, {\rm ergs} \, {\rm
s}^{-1}$) exceeds by a few orders of magnitude the value derived
for magnetospheric optical emission of young pulsars (Pacini
\cite{pa}). This fact together with the negative result of
searching of optical pulsations at the radio period led to the
conclusion that the proposed identification could be erroneous
(\cite{mi}; see also \cite{she}, \cite{chak}). We suggest,
however, that the observed optical emission should rather be
attributed to the bow shock around the pulsar than to the pulsar
itself. This suggestion is supported by the estimate of the total
luminosity of the bow shock, $L_{\rm T} \simeq Sn_{\rm cl} v_\ast
(m_{\rm p} v_\ast ^2 /2)$, where $S\simeq \pi r_{\rm b} ^2$ and
$r_{\rm b} \simeq 3^{1/2} r_{\rm s}$ are the area and the
characteristic radius of the bow shock, respectively. For the
adopted parameters\footnote{Note that shocks in the range ($v_{\rm
s} \geq 40\, {\rm km}\,{\rm s}^{-1}$, where $v_{\rm s}$ is the
shock velocity ($\simeq v_\ast$), $10^6 \, {\rm cm}^{-3} \leq
n_{\rm cl} \leq 10^8 \, {\rm cm}^{-3}$) produce strong 22\,GHz
${\rm H}_2 {\rm O}$ masers (see Fig. 3 of \cite{h}). The possible
detection of ${\rm H}_2 {\rm O}$ masers towards PSR B\,1509-58
will lend strong support to the suggestions of this paper.} this
gives $L_{\rm T} \simeq \alpha \cdot 10^{33} \, {\rm ergs} \, {\rm
s}^{-1}$, i.e. $L_{\rm V} \simeq L_{\rm T}$ for $\alpha \simeq 1$.
It is obvious that if our suggestion is correct, there can be no
correspondence between the observed luminosity and the luminosity
expected from the results of Pacini (\cite{pa}). The optical
pulsations should be absent as well.

To conclude, we point out a curious coincidence of the accretion
rate derived in Sect. 2 with accretion rates required in
accretion-based models to explain high spin-down rates of
anomalous X-ray pulsars and soft gamma-ray repeaters (e.g.
\cite{me}, \cite{gho}, \cite{cha}, \cite{a}). This coincidence
allows us to believe (\cite{g9a}, 2000\cite{g0b}) that these
objects could lose a significant part of their rotational energy
due to the process discussed in this paper, and that their ``true"
ages could be much larger than the respective characteristic
spin-down ones.

\begin{acknowledgements}
I am grateful to N.D'Amico and A.D'Ercole for discussions, to D.Page
(the referee) for comments, and to J.K.Katgert-Merkelijn (the Deputy
Editor) for carefully reading the manuscript.
This work was partially supported by NPS.
\end{acknowledgements}

\end{document}